\documentclass[journal]{IEEEtran}

\usepackage{xcolor}
\usepackage{amsmath}
\usepackage{algpseudocode}
\usepackage{algorithm}
\usepackage[font=footnotesize]{caption}
\usepackage{multicol}
\usepackage{multirow}
\usepackage{cite}
\usepackage{comment}
\usepackage{makecell}
\ifCLASSINFOpdf
\else
\fi

\usepackage{tikz,xcolor,hyperref}

\definecolor{lime}{HTML}{A6CE39}
\DeclareRobustCommand{\orcidicon}{%
	\begin{tikzpicture}
	\draw[lime, fill=lime] (0,0) 
	circle [radius=0.16] 
	node[white] {{\fontfamily{qag}\selectfont \tiny ID}};
	\draw[white, fill=white] (-0.0625,0.095) 
	circle [radius=0.007];
	\end{tikzpicture}
	\hspace{-2mm}
}
\foreach \x in {A, ..., Z}{%
	\expandafter\xdef\csname orcid\x\endcsname{\noexpand\href{https://orcid.org/\csname orcidauthor\x\endcsname}{\noexpand\orcidicon}}
}

\begin{document}

\title{An Enhanced Energy Management System Including a Real-Time Load-Redistribution Threat Analysis Tool and Cyber-Physical SCED}

\author{
Ramin~Kaviani\orcidA{}, \IEEEmembership{Student Member,~IEEE,}
Kory W.~Hedman\orcidB{},~\IEEEmembership{Senior Member,~IEEE}
\thanks{This work has been implemented to fulfill a part of the project: “A Verifiable
Framework for Cyber-Physical Attacks and Countermeasures in a Resilient
Electric Power Grid” funded by the National Science Foundation (NSF) Award
under Grant 1449080.} 
}

%\markboth{IEEE TRANSACTIONS ON POWER SYSTEM, VOL., NO., 2020}%

\maketitle

\begin{abstract}
It is possible to launch undetectable load-redistribution (LR) attacks against power systems, even in systems with protection schemes. Therefore, detecting LR attacks in power systems and establishing a corrective action to provide secured operating points are imperative. In this paper, we develop a systematic real-time LR threat analysis (RTLRTA) tool, which can flag LR attacks and identify all affected transmission assets. Since attackers might use random deviations to create LR attacks, we introduce an optimization model to generate random LR attacks. Hence, we can determine accurate thresholds for our detection index and test the tool's functionality when there are random LR attacks. 
Additionally, based on an estimation for the actual loads in the post-attack stage, we design a set of physical line flow security constraints (PLFSCs) and add it to the security-constrained economic dispatch (SCED) model. We call the new model cyber-physical SCED (CPSCED), which can appropriately respond to the identified LR attacks and provide secured dispatch points. We generate multiple scenarios of random LR attacks and noise errors for different target lines in the $2383$-bus Polish test system to validate our proposed methods' accuracy and functionality in detecting LR attacks and responding to them.
\end{abstract}


\IEEEpeerreviewmaketitle

\section*{Nomenclature}
	\addcontentsline{toc}{section}{Nomenclature}
	\emph{\textbf{Sets and Indices}}
    \begin{IEEEdescription}[\IEEEsetlabelwidth{The   Label}] 
  \item [{$G$}] Set of all generation units.
  \item [{$g$}] Index for generation unit.
  \item [{$G(i)$}] Set of all generation units at bus $i \in N$.
  \item [{$i$}] Index for bus.
  \item [{$K$}] Set of all transmission assets.
  \item [{$K^T$}] Set of all affected transmission assets by an LR attack; $K^T \subset K$.
  \item [{$k$}] Index for transmission asset.
%  \item [{$M$}] Set of all measurements.
%  \item [{$m$}] Index for measurement.
  \item [{$N$}] Set of all buses.
  \item [{$\eta^k$}] A randomly selected set of buses from all sensitive buses with respect to a target asset $k \in K$.
   \item [{$\psi$}] Set of all buses with load deviations more than $\alpha_{min}^{startpoint}$ $\times$ forecasted loads; $\psi \subset N$.
  \end{IEEEdescription}

\emph{\textbf{Variables}}
   \begin{IEEEdescription}[\IEEEsetlabelwidth{The  Label}] 
  \item [{$\mathbf{H'_i  \Delta\theta}$}] Load deviation at bus $i \in N$ (MW).
%  \item [{$\hat{x}$}] $n_b \times 1$ vector of estimated state variables.
%  \item [{$c$}] $ n_b \times 1 $ vector of false data generated by attacker.
  \item [{$P_g$}] Dispatch point of unit $g \in G$ (MW).
  \item [{$P_k$}] Active power flow on transmission asset $k \in K$ (MW) considering the loads from SE.
  \item [{$PLF_k$}] Active power flow on transmission asset $k \in K^T$ (MW) considering the actual loads.
  \item [{$D_i^a$}] Actual load's estimation (MW) at bus $i \in N$.
%  \item [{$x$}] $n_b \times 1$ vector of actual state variables.
%  \item [{$\eta$}] $n_b \times 1$ vector of false data generated by attacker.
  \item [{$P_{inj}^i$}] Active power injection (MW) to bus $i \in N$ considering the loads from SE.
  \item [{$P_{inj,i}^*$}] Active power injection (MW) to bus $i \in N$ considering the actual loads.
  \item [{$\mathbf{\Delta\theta}$}] Vector of Buses' angle deviation.
  \end{IEEEdescription}

\emph{\textbf{Parameters, Vectors and Matrices}}
    \begin{IEEEdescription}[\IEEEsetlabelwidth{The Label}]
 % \item [{$n_b$}] Number of buses.
  \item [{$\alpha$}] Load shift factor.
  \item [{$\alpha_{k, min}^{start point}$}] Minimum load shift factor that causes transmission asset $k \in K$ to have overflow.
  \item [{$\alpha_{k, min}^{5 \%}$}] Minimum load shift factor that causes transmission asset $k \in K$ to have $5 \%$ overflow.
%  \item [{$\Bar{P}_g$}] Fixed dispatch point of unit $g \in G$.
%  \item [{$\tau$}] Residual-based bad data detector threshold.
  \item [{$c_g$}] Production cost of unit $g \in G$.
%  \item [{$e$}] $n_m \times 1$ vector of measurement noise errors.
%  \item [{$H$}] $n_m \times n_b$  Jacobian matrix of the system.
  \item [{$\mathbf{H^{'}}$}] $n_b \times n_b$ dependency matrix between power injection measurements and state variables.
  \item [{$\mathbf{H_i^{'}}$}] $i^{th}$ row of $H^{'}$ ($i \in N$).
  \item [{$L_i$}] The contaminated load (MW) at bus $i \in N$ resulted from SE.
  \item [{$\mathbf{L^{it}}$}] The vector of contaminated loads (MW) resulted from SE at iteration $it$ of the EMS.
%  \item [{$lb_i$}] Lower bound for load deviation at each bus $i \in N$.
%  \item [{$N_1$}] Number of states that can be changed by attacker.
%  \item [{$n_m$}] Number of measurements.
%  \item [{$n_{br}$}] Number of branches.
  \item [{$P_g^{max}$}] Upper limit (MW) on capacity of unit $g \in G$.
  \item [{$P_g^{min}$}] Lower limit (MW) on capacity of unit $g \in G$.
  \item [{$P_k^{max}$}] Continuous thermal rating (MW) of transmission asset $k \in K$.
  \item [{$PTDF_{k,i}^R$}] Power transfer distribution factor for transmission asset $k \in K$ and bus $i \in N$ (injection) with regard to reference bus R (withdrawal).
%  \item [{$ub_i$}] Upper bound for load deviation at each bus $i \in N$.
%  \item [{$Y$}] $ n_m \times 1 $ vector of measurements.
%  \item [{$W$}] Diagonal matrix of measurements' weight.
%  \item [{$\sigma_m$}] Standard deviation of error introduced to each measurement $m \in M$.
% \item [{$\mathbf{\Delta L^{it}}$}] Vector of Load deviations at iteration $it$ of the EMS.
  \item [{$\mathbf{D^{it}}$}] Vector of forecasted loads (MW) at iteration $it$ of the EMS.
  \item [{$D_{i}$}] Forecasted load (MW) at bus $i \in N$.
  \item [{$d$}] Number of sensitive buses that are randomly selected to be zero in random LR attacks; $d = |\eta|$.
%  \item [{NPDSB$_k$}] Best location deviation magnitude and direction index for target line $k \in K$.
% \item [{NPDSB$_^{ref}_k$}] Threshold value for NPDSB of line $k \in K$.
  \end{IEEEdescription}

\section{Introduction}
\IEEEPARstart{D}{ue} to the extensive usage of cyber layers to monitor, control, and optimize the real-time operations of power systems, many research studies have addressed the challenges associated with these cyber layers and the risk of cyber-attacks. The research studies concerning cyber-attacks are divided into two parts: 1) implementing and modeling cyber-attacks, and 2) designing protection, detection, and corrective schemes against cyber-attacks.
\subsection{Implementing Cyber-Attacks}
In the literature, one of the popular ways to generate cyber-attacks against power systems is to create a false data injection attack (FDIA). 
In FDIAs, attackers add malicious data to the actual measurements in such a way that they can bypass the existing residual-based bad data detectors and achieve the desired state estimation (SE)'s output \cite{liu2011false}. FDIAs could be created based on different goals, such as overloading a transmission asset \cite{liang2016vulnerability,chakrabarty2020detection,chu2016evaluating}, changing the actual topology of a system \cite{zhang2016physical},  changing locational marginal prices (LMPs) \cite{xie2011integrity,tajer2017false}, increasing the operational cost/loss of a system \cite{yuan2011modeling,liu2016cyber}, causing sequential outages \cite{che2018false}, and  frequency instability \cite{ameli2018attack,huang2018online}. Likewise, there are different ways to implement FDIAs based on the measurements that should be compromised. For instance, the authors in \cite{zhang2016physical} attempted to falsify the transmission lines status' information and measurements to perform topology-based FDIAs. In \cite{chakrabarty2020detection}, the authors falsified the phase shift commands to launch a transmission asset overloading based FDIA. The authors in \cite{ameli2018attack,huang2018online} falsified the automatic generation control (AGC) signals to attack the systems' stability. The authors in \cite{tan2017cyber,liu2014local,liu2015modeling, kaviani2019identifying} changed the load and power flow measurements to perform load-redistribution (LR) based FDIAs. This paper focuses on LR attacks, which falsify the load measurements to maximize the physical overflow on a target transmission asset.

In LR attacks, the attackers falsify the buses' injection measurement by changing the load measurement at each bus; they avoid falsifying the generation at each bus and injection measurements at zero injection buses (attempt not to increase the detection risk). In \cite{yuan2011modeling,liu2016cyber}, the authors attempted to maximize the operations' costs by designing bi-level LR attack models. In \cite{chu2016evaluating,liang2016vulnerability}, the authors modeled two different bi-level LR attacks to maximize the power flows of critical transmission assets while penalizing the number of resources that attackers can access. In \cite{tan2017cyber}, the authors proposed a bi-level mixed-integer linear programming LR attack model to overload multiple transmission assets. In \cite{liu2014local,liu2015modeling}, the authors designed LR attacks with incomplete systems' information by proposing a model to find the best local attacking region.

\subsection{Countermeasures against Cyber-Attacks}
Due to the catastrophic consequences that cyber-attacks could have for power systems, the research community has been pushed to seek a solution and develop countermeasures against cyber-attacks. In power systems, the security actions to stand against cyber-attacks are divided as follows:

\subsubsection{\textbf{Protection Schemes}}
Refer to all actions that are done in pre-attack stages to make systems secure against cyber-attacks. In other words, these actions are designed to prevent attackers from being able to launch cyber-attacks against power systems. The authors in \cite{yang2017pmu,deng2017defending, liang2018distributed, amini2016dynamic, sreeram2019managing, wei2016stochastic, liu2020pre} proposed various protection techniques, such as a blockchain-based framework to decentralize the data managing systems, stochastic game theory models to find the optimal way of protecting critical elements, and greedy algorithms to find protection strategies, which minimize the systems' vulnerabilities.

\subsubsection{\textbf{Detection Schemes}}
Most of the protection techniques are expensive or cause significant disruptions in the systems' infrastructure. Moreover, the research studies in \cite{sreeram2019managing,deka2016jamming} demonstrated the attackers' ability to launch cyber-attacks even in systems with one insecure measurement. These facts have pushed the researchers to design and develop static/dynamic based detection mechanisms to continue the process of standing against cyber-attacks \cite{ pan2019static,chakhchoukh2019diagnosis,chakrabarty2020detection,zhao2016robust,gao2016identification,liu2014detecting,pal2017classification,kaviani2019detection, esmalifalak2017detecting,ozay2016machine,foroutan2017detection,zhang2020detecting}. 

\subsubsection{\textbf{Corrective Actions}} Even after the attacks are identified, retrieving the affected measurements' actual value may be impossible for operators. It is then imperative for system operators to take corrective actions to mitigate the attacks' physical consequences and avoid any severe damage (e.g., cascading outages). In this regard, the studies in \cite{che2018mitigating,shayan2019network} addressed some post-attack corrective actions to provide secured operating points.

\subsection{Our Contributions}
This paper's primary goals are 1) developing a systematic tool for flagging LR attacks in real-time and identifying the affected transmission assets (if exist) and 2) designing a corrective action to respond to the identified attacks and provide secured dispatch points.

For the detection part, we use the security index proposed in \cite{kaviani2019detection}, the number of proper deviations at sensitive buses (NPDSB), to detect LR attacks, and based on that, develop a real-time load-redistribution threat analysis (RTLRTA) tool. However, we suggest a more accurate way to determine the thresholds for different NPDSB indices associated with different target transmission assets. Our proposed detection mechanism is different from other proposed techniques in the literature in various aspects, like:
\begin{itemize}
\item As opposed to \cite{pan2019static,chakhchoukh2019diagnosis}, in which the authors developed dynamic-based detection mechanisms, our detection mechanism is a static-based.
\item As opposed to the assumption in \cite{chakrabarty2020detection}, our method successfully detects LR attacks assuming that attackers have no limitation for altering state variables. As opposed to the approach in \cite{zhao2016robust}, our mechanism does not rely on some secured measurements.
\item As opposed to the proposed techniques in \cite{gao2016identification,liu2014detecting}, our detection mechanism is modeled based on a linear and convex problem.
\item As opposed to the proposed method in \cite{pal2017classification}, in which the method's functionality was not evaluated in the presence of normal noise errors, we investigate our detection mechanism's functionality in the presence of both Gaussian and non-Gaussian noise errors.
\item As opposed to the methods in \cite{esmalifalak2017detecting,ozay2016machine,foroutan2017detection,zhang2020detecting}, which were developed based on machine and deep learning, our mechanism perfectly works regardless of the available historical data's quantity and quality (learning-based methods might be more straightforward and effective, but need a large amount of underlying historical data).  
\end{itemize}

For developing the corrective action, we add a set of physical line flow security constraints (PLFSCs) to the security-constrained economic dispatch (SCED) model and introduce the cyber-physical SCED (CPSCED). The CPSCED model provides secured dispatch points concerning the identified attacks and affected transmission assets from the RTLRTA tool. As opposed to the proposed corrective actions in \cite{che2018mitigating,shayan2019network}, which were developed based on complicated and time-consuming tri-level optimization problems, our proposed remedial  action is swift and straightforward. 
In brief, our main contributions are:
\begin{enumerate}
\item Improving the accuracy of the threshold proposed in \cite{kaviani2019detection} and introducing  a new approach to determine thresholds for NPDSB indices.

\item Introducing and developing the fast RTLRTA tool, with minimal disruptions in the existing EMSs, to detect any possible LR attack and identify all affected transmission assets.

\item Introducing and designing the straightforward CPSCED model, yet, practically efficient for the real-time operations. The CPSCED does a real-time corrective action that allows operators to remove or mitigate the identified attacks' physical consequences.

\item Because all security actions should be addressed in real-time with minimal changes and disruptions in the existing EMSs’ infrastructure, we develop our detection and corrective schemes highly effective, fast, and applicable to real-world practice. 
\end{enumerate}

The rest of this paper is organized as follows. Section II presents a brief background regarding the SCED model and LR attacks. Section III introduces the new enhanced EMS, including the RTLRTA tool and the CPSCED. Sections IV and V illustrate the simulation results and concluding remarks, respectively.
% === II. Background ========================================
\section{Background}
\subsection{Security-Constrained Economic Dispatch (SCED)}\label{A}
In the real-time operations of power systems, energy is cleared through an economic dispatch model. 
This section shows a simple base-case SCED model, which considers all the grid's and units' physical limitations in the pre-contingency stage (this model does not include reserve requirements and the post-contingency security constraints). 
\begin{flalign}
\max_{{P_g}} \quad 
&\sum_{g \in G} c_gP_g \label{SCED-OBJ}, \\
\text{s.t.}\quad
& \sum_{g \in G} P_g = \sum_{i \in N} L_i, \label{POW-BAL} \\
& P_k = \sum_{i \in N} PTDF_{k,i}^R P_{inj}^i; ~ \forall k \in K \label{LIN-FLOW}, \\
& P_{inj}^i = \sum_{g \in G(i)} P_g - L_i; \; \forall i \in N \label{POW-INJ}, \\
& -P_k^{max} \leq P_k \leq P_k^{max}; \; \forall k \in K \label{LIN-LIM}, \\
& P_g^{min} \leq P_g \leq P_g^{max}; ~ \forall g \in G \label{GEN-CAP}.
\end{flalign} 

The objective function in (\ref{SCED-OBJ}) minimizes the production cost of the power needed to meet the demand. The procurement of enough energy to meet the demand is imposed in (\ref{POW-BAL}). The DC approximation of each transmission asset's active power flow is formulated using the power transfer distribution factors (PTDFs) in (\ref{LIN-FLOW}). The power injected to each bus and transmission network's limitations are modeled in (\ref{POW-INJ}) and (\ref{LIN-LIM}), respectively. In (\ref{GEN-CAP}), the generation units' physical limitations are modeled.
\subsection{LR Attack}\label{B}
LR attack is a type of FDIAs, in which changing buses' injection measurement is the attackers' procedure to achieve their goals. In this paper, due to the direct communication between the generation units' control room and system operators, the only way to change the buses' injection measurement is to change the buses' load measurement. In brief, an LR attack increases the loads at some buses and decreases other buses' load subject to the attacker's limitations. The load deviation at each bus should be neither more nor less than a fixed fraction of the forecasted load at that bus. Moreover, the total load in the system has to remain unchanged.

This paper focuses on the LR attacks that are designed to cause an overflow on a target transmission asset. We use the special structure of LR attacks' core problem, demonstrated in (\ref{Algorithm_3's_Alternative_1})-(\ref{Algorithm_3's_Alternative_3}), as the LR attack model throughout this study. 

\begin{flalign} 
\max_{{\mathbf{H'_i  \Delta \theta}}} \quad
& \pm\sum_{i \in N} (\mathbf{H'_i   \Delta \theta})PTDF_{l,i}^{R}\label{Algorithm_3's_Alternative_1}, \\
\text{s.t.}\quad
& -\alpha D_i \leq (\mathbf{H'_i   \Delta \theta}) \leq \alpha D_i, \label{Algorithm_3's_Alternative_2} \\
& \sum_{i \in N} (\mathbf{H'_i   \Delta \theta}) = 0 \label{Algorithm_3's_Alternative_3}.
\end{flalign}

In this model, `$\pm$' indicates that the load deviations' directions depend on the target asset's initial flow direction (might be positive or negative). The deviations' directions for a target asset with a positive initial flow direction are different from the deviations' directions associated with a target asset with a negative initial flow direction. The primary decision variable is $\mathbf{H'_i   \Delta \theta}$, which indicates the net injection deviation at bus $i \in N$ (we used $\mathbf{H'_i   \Delta \theta}$ to emphasize that attackers can change bus angles to get appropriate deviations in loads). The load shift factor is shown by $\alpha$, and $D_i$ denotes the forecasted load at each bus $i \in N$. 
The power transfer distribution factor of the target asset $l \in K$, with respect to the injection at bus $i \in N$ and withdrawal from the reference bus $R$, is shown by $PTDF_{l,i}^{R}$.

In this problem, the objective function maximizes the overflow on a target transmission asset. Constraints in (\ref{Algorithm_3's_Alternative_2}) limit the attackers from changing the load at each bus more/less than $+$/$-\alpha$ percent of the forecasted load at that bus (they also impose no change at zero injection buses). Constraint (\ref{Algorithm_3's_Alternative_3}) ensures that the system's net load remains unchanged. 

% === III. Modeling and Methodology ========================================
\section{Modeling and Methodology}
\begin{comment}
In the first part of this section, we developed the RTLRTA tool based on the NPDSB index, proposed in \cite{kaviani2019detection}, but using a more accurate way to determine thresholds for NPDSB indices. In the second part, we designed the CPSCED model by adding PLFSCs to the SCED model.
\end{comment}
In the first part of this section, we go through the process of developing the RTLRTA tool using the NPDSB security index. In the second part, we introduce a way to estimate the actual loads after LR attacks. Then, using the actual loads' estimation, we design the PLFSCs and go through the process of CPSCED modeling.

\subsection{The RTLRTA Tool}\label{RTLRTA}

\subsubsection{NPDSB}\label{NPDSB}
In power systems, KVL and KCL govern the power flows on transmission assets. According to this fact, the only way for attackers who want to change the loads to achieve the maximum overflow on a target transmission asset is to have load deviations with proper directions and magnitudes at buses with the largest PTDFs. 
Considering this fact, the authors in \cite{kaviani2019detection} proposed the NPDSB index, which shows the number of proper deviations at sensitive buses associated with a set of loads and a target asset. Then, if the index's value related to a set of loads is greater than a threshold, which in \cite{kaviani2019detection} was assumed to be half of the total number of sensitive buses, that set of loads is flagged as a malicious set. 

This paper propose a more accurate procedure to determine thresholds for NPDSB indices since the proposed threshold in \cite{kaviani2019detection} may not be accurate enough to detect all random LR attacks and distinguish them from random noise errors. 
In other words, there is no unique threshold for all NPDSB indices; instead, there are different thresholds associated with different NPDSB indices.     

\subsubsection{Thresholds for NPDSB Indices Considering Random LR Attacks}
In this subsection, we propose a procedure to find more accurate thresholds for NPDSB indices. To do so, at first, we re-design problem (\ref{Algorithm_3's_Alternative_1})-(\ref{Algorithm_3's_Alternative_3}) and model problem (\ref{oldattackmodel})-(\ref{newattackmodel_1}) to generate random LR attacks.
\begin{flalign}
& (\ref{Algorithm_3's_Alternative_1})-(\ref{Algorithm_3's_Alternative_3}) \label{oldattackmodel}, \\
& \mathbf{H'_d \Delta\theta} = 0; ~ \forall  d  \in \eta^k. \label{newattackmodel_1}
\end{flalign}

We add constraint (\ref{newattackmodel_1}) to force the deviations at $d$ randomly selected sensitive buses (concerning the target transmission asset) to be zero.

The threshold values should be determined in a way that they can detect even the weakest random LR attacks. Therefore, we can consider the NPDSB index of the weakest, yet effective, random LR attack against a target transmission asset as the threshold for that asset's NPDSB index. Due to the inverse relationship between the NPDSB index and $d$, we can find the thresholds by solving problem (\ref{oldattackmodel})-(\ref{newattackmodel_1}) with the largest value of $d$. We provide more clarifications and detailed information about the process of finding thresholds for the NPDSBs of different transmission assets in section \ref{results}.

\subsubsection{Developing The RTLRTA Tool}\label{RTLRTA_sub}
Here, we leverage the NPDSB index to develop the RTLRTA tool to detect LR attacks and find all affected transmission assets in real-time.  The RTLRTA tool has the same inputs as SE and calculates the NPDSB indices associated with all or only vulnerable assets. Then, the RTLRTA tool compares the resulted NPDSB indices with the pre-determined thresholds to find out whether the current set of loads has been contaminated with malicious data or not.

One of the RTLRTA tool’s advantages is that it could identify all affected transmission assets. There might be correlations between critical transmission assets in a power system, so a set of malicious load deviations, designed initially to damage a specific transmission asset, may affect other transmission assets. For instance, assume a power system with five vulnerable transmission lines A, B, C, D, and E. For this system, a random LR attack scenario against line C might exist that can affect lines B and E. Likewise, another attack scenario might exist against this line, which can affect lines A and E. Based on this fact, we develop the RTLRTA tool to check the NPDSB indices for all vulnerable transmission assets and finds all affected ones (if any exists). As a result, the RTLRTA tool's outputs are the affected transmission assets' ID or index. Therefore, the RTLRTA not only determines whether a system is under an LR attack, but it also identifies the affected transmission assets. The output from this tool is then fed into the CPSCED to activate the corresponding PLFSCs.

Algorithm \ref{RTLRTANPDSB} demonstrates the process of calculating the NPDSB index for each transmission asset ($\alpha_{k,min}^{start point}$ is the smallest load shift factor that causes an overflow on the target transmission asset $k$). Moreover, considering the NPDSB indices and their associated thresholds, Algorithm \ref{RTLRTANPDSB} finds the affected assets' ID (to send them to the CPSCED for PLFSCs activation).

\begin{algorithm}[!h]
\scriptsize
\caption{Process of finding NPDSB indices and  affected transmission assets in RTLRTA} 
\textbf{Input:} Output from SE.\\
\textbf{Output:} NPDSB indices and affected assets' ID.
\begin{algorithmic}[1]
\For{\texttt{$it$ $\gets$ EMS~ iteration}}
\State  $\Delta L^{it}\gets  L^{it} - D^{it}$;
\For{\texttt{$k \gets K$}}
\State NPDSB$_k$ $\gets 0$;
\State  $[H'_i\Delta\theta]^k$ $\gets$ solve ~problem~ (\ref{Algorithm_3's_Alternative_1})-(\ref{Algorithm_3's_Alternative_3});
\For{\texttt{$i \gets N$}}
\If{($sign[\Delta L_i^{it}] = sign[H'_i \Delta\theta]_i^k$  \& $|\Delta L_i^{it}| \geq \alpha_{k,min}^{start point}D_i~)$}
\State  NPDSB$_k$ $\gets$ NPDSB$_{k} + 1$;
\EndIf
\EndFor

\If { NPDSB$_k$ $ \geq$ NPDSB$_{threshold}^k$}
\State PLFSC$_k$ is activated;
\EndIf

\EndFor
\EndFor

\end{algorithmic}
\label{RTLRTANPDSB}
\end{algorithm}

\subsection{Cyber-Physical SCED (CPSCED)}\label{CPSCED}
The rationale behind using the CPSCED is that after noticing an LR attack in the system, it may still be hard to achieve the actual loads. Therefore, considering the fake load measurements, a fast corrective action should be taken to maintain the system’s operation secure. Subsequently, we modify the SCED model by adding the PLFSCs to create the CPSCED model, which provides secured dispatch points and avoids physical overflows or mitigates significant impacts.

Assume an attacker bypasses all protection-based schemes and changes the buses’ load. As a result, the set of insecure dispatch points from the SCED is $P^{*}_g$, which creates physical overflows in the system (considering the actual loads). According to this strategy, we embed two sets of line flow security constraints in the CPSCED. The first one, which already exists in the base-case SCED, imposes the power flows on transmission assets within their ranges (considering the contaminated loads). The second one, the new PLFSC, forces the physical flows on affected transmission assets to be within their boundaries (considering the actual loads). However, because retrieving the actual loads is hard, we propose a method to estimate the actual loads, then we modeled the PLFSC based on this estimation.

To estimate the actual loads, we assume the worst-case LR attack for the affected transmission asset with the largest NPDSB index value and model the actual load at each bus as follow:

\begin{equation}
    D_i^a = \begin{cases} 
           L_i \pm  \mathbf{H'_i \Delta\theta}, ~~~ \text{if} ~ i \in \psi\\
          L_i, ~~~~~~~~~~~~~~ \text{if} ~ i \notin \psi
            \end{cases}
\end{equation}
where negative ($L_i -  \mathbf{H'_i \Delta\theta}$) is used for transmission assets with positive initial flow directions and positive ($L_i +  \mathbf{H'_i \Delta\theta}$) is used for transmission assets with negative initial flow directions.
Additionally, since the attacker might introduce some random deviations, the actual load at each bus $i \in N$ with $\Delta L_i$ more than $\alpha^{startpoint}_{k,min} \times D_i$ is modeled as $ L_i \pm  \mathbf{H'_i \Delta\theta}$; otherwise, it remains $L_i$. 

After estimating the actual loads, we can model the PLFSC, as shown in  (\ref{CP-PLFSC})-(\ref{CP-TAR-LIN-LIM}). Then, by adding the PLFSC to the SCED model, we can design the CPSCED as follow: 
\begin{flalign}
& (\ref{SCED-OBJ})-(\ref{GEN-CAP}) \label{SCED_model},\\
& PLF_k = \sum_{i \in N} SF_{k,i}^R P_{inj,i}^*; ~ \forall k \in K^T \label{CP-PLFSC}, \\
& P_{inj,i}^* = \sum_{g \in G(i)} P_g - D_i^a; ~ \forall i \in N \label{CP-PLFSC-INJ}, \\
& -P_k^{max} \leq PLF_k \leq P_k^{max}; ~ \forall k \in K^T \label{CP-TAR-LIN-LIM}.
\end{flalign}
where in (\ref{CP-PLFSC}), the actual physical flow is formulated based on the actual injection to each bus, which is modeled in (\ref{CP-PLFSC-INJ}). Constraint (\ref{CP-TAR-LIN-LIM}) forces the physical flow of each affected transmission asset to be within its thermal limits.

\begin{figure}[!h]
 \centering
 \includegraphics[trim = 1mm 37mm 5mm 3mm, clip, width=9.9cm]{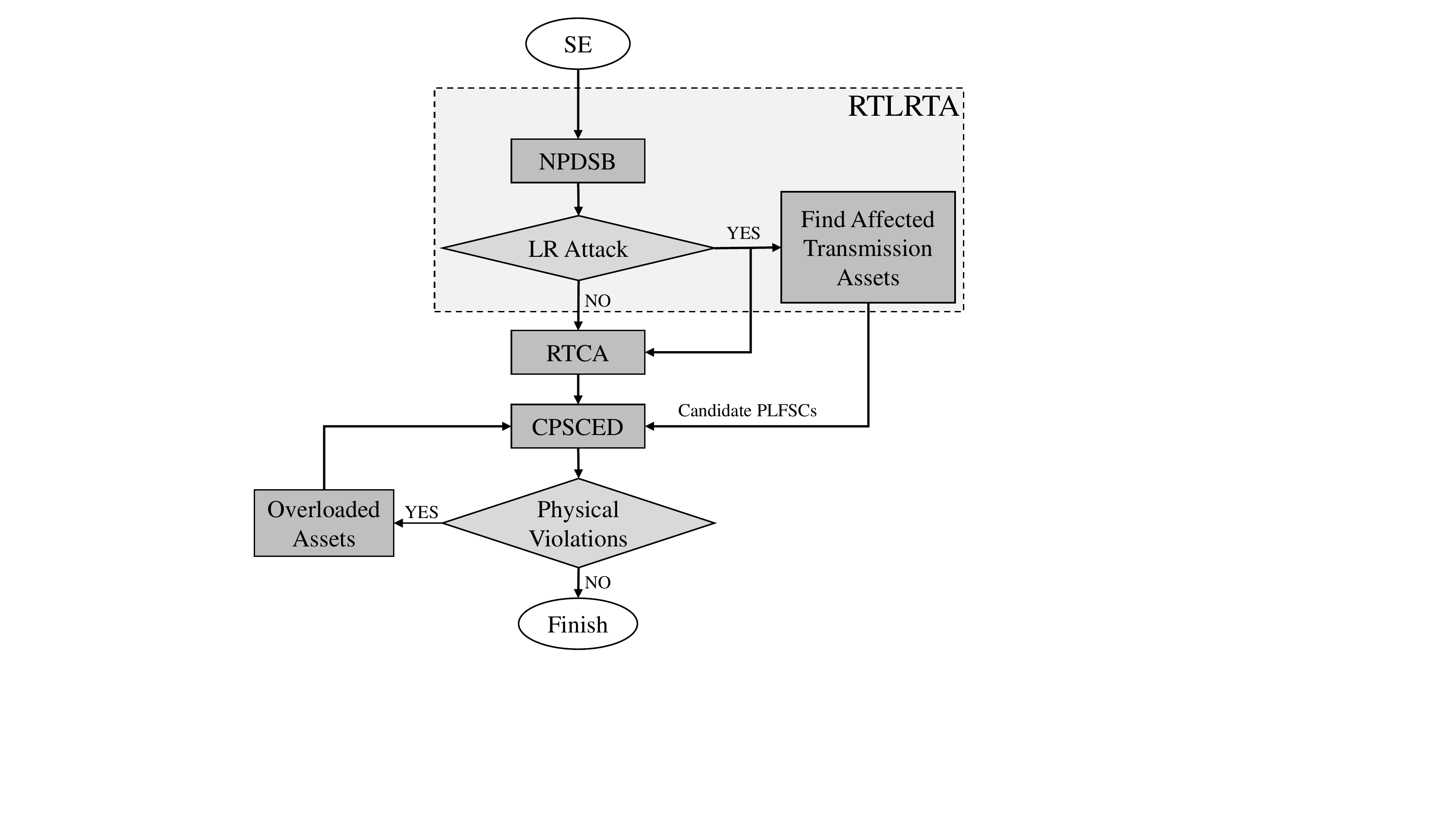}
 \caption{Enhanced EMS including the RTLRTA tool and CPSCED.}
 \label{enhanced-EMS}
\end{figure}

The proposed enhanced EMS algorithm is designed and shown in Algorithm \ref{enhancedEMS}, and the block diagram of the enhanced EMS is shown in Fig. \ref{enhanced-EMS}.

\begin{algorithm}[!h]
\scriptsize
\caption{The Enhanced EMS algorithm.}
\begin{algorithmic}[1]
\State Get the SE's results;
\State  In the RTLRTA tool, find NPDSBs$_{Threshold}$ for vulnerable transmission assets, considering $\alpha$ at most $10 \%$;
\State  In the RTLRTA tool, flag possible LR attacks and identify all affected transmission assets based on the NPDSBs$_{Threshold}$;
\State  In the CPSCED, activate PLFSCs corresponding to identified affected assets;
\State  Run the CPSCED based on the estimated actual loads ($L^{it}_i \pm \mathbf{H'_i  c}$);
\State  Get the new dispatch points and find the actual physical flows;   
\State If there is overflow, add the corresponding PLFSCs to the CPSCED;
\State Go to step $5$;
\end{algorithmic}
\label{enhancedEMS}
\end{algorithm}

At each iteration, the NPDSB indices for all vulnerable assets are calculated. Then, if there are affected assets, the CPSCED is solved with the corresponding activated PLFSCs. Moreover, considering the CPSCED's dispatch points and the estimated actual loads, other transmission assets' physical flows are checked. If overloaded transmission assets still exist, the corresponding PLFSCs are activated in CPSCED, and a new set of dispatch points is provided. This process continues until there is no physical overflow in the system.

% === IV. simulation and results ========================================

\section{Simulation Results}\label{results}
In this section, we applied the proposed methods to the $2383$-bus test system\cite{zimmerman2011matpower}. At first, we determined the threshold values for two vulnerable lines $169$ and $251$ (they are vulnerable lines since there is at least one scenario of LR attack for each line with $\alpha$ at most $10 \%$ that can make the line physically overloaded). Then, we evaluated the RTLRTA tool's and CPSCED's accuracy and functionality for different random LR attacks. At last, we compared the results from the CPSCED with the results from the SCED.  

\subsection{NPDSB$^{169}_{Threshold}$}\label{thresh169}
Here, we determined a threshold for the NPDSB index of transmission line $169$. To do so, we considered the weakest LR attack that could cause an overflow on this line. Since the largest $d$ is important to find the weakest attack with the lowest NPDSB index, we started to force the deviations at the first $d$ sensitive buses out of all sensitive buses to zero. The sensitive buses are sorted based on their PTDF absolute values from smallest to largest. 
Throughout this experiment, we increased $d$ manually and solved problem (\ref{oldattackmodel})-(\ref{newattackmodel_1}) until we reached a $d$ whose associated attack vector could not cause an overflow on line $169$. Table \ref{line169thresh} shows the results of solving problem (\ref{oldattackmodel})-(\ref{newattackmodel_1}) for three different $d$s.  

As shown in Table \ref{line169thresh}, there is no random LR attack with $\alpha$ at most $10 \%$ and $d$ more than $692$ (NPDSB $\leq$ $357$) that could cause an overflow on line $169$. It means that the system operator can use $357$ as NPDSB$^{169}_{threshold}$. We used $350$ for this line, which is even more conservative than $357$.

\begin{table}[!h]
\scriptsize
\centering
\caption{The control room flows, physical flows, and NPDSB indices after solving problem (\ref{oldattackmodel})-(\ref{newattackmodel_1}) for line $169$ with $\alpha = 10 \%$, $\alpha^{startpoint}_{169, min} = 0.0425$,  and three different $d$s.}

\begin{tabular}{|c|c|c|c|c|c|} \hline
\multirow{2}{*}{\makecell{\textbf{Line}}} & \multirow{2}{*}{\makecell{\textbf{d}}} &  \textbf{Control room} & \textbf{Physical} &  \multirow{2}{*}{\makecell{\textbf{NPDSB}}} & \textbf{Flow limit}   \\
&  &  \textbf{flow (MW)} & \textbf{flow (MW)} & & \textbf{(MW)} \\
\hline
\multirow{3}{*}{\makecell{169}}  & 292 &  -926.62 & -1166.68 & 667 & \multirow{3}{*}{\makecell{926.62}} \\ \cline{2-5}
 &492 &  -916.35 & -1132.68 & 526 &  \\ \cline{2-5}
 &  692 &  -765.76& -926.65 & \textbf{357} &  \\
\hline
\end{tabular}
\label{line169thresh}
\end{table}

\subsection{NPDSB$_{Threshold}^{251}$}

Following the same procedure in subsection \ref{thresh169}, we determined a threshold for the NPDSB index of line $251$.

Table \ref{line251thresh} demonstrates the control room flows, physical flows and NPDSB indices corresponding to three different $d$s for line $251$ with $\alpha$ and $\alpha^{startpoint}_{251, min}$ equal to $10 \%$ and $0.0686$, respectively. According to Table \ref{line251thresh}, there is no severe random LR attack for line $251$ with $\alpha \leq 10 \%$ and $d \geq 685$ (NPDSB $\leq 374$). Therefore, $370$ is a valid and conservative enough choice for NPDSB$_{threshold}^{251}$. 

\begin{table}[!h]
\scriptsize
\centering
\caption{The control room flows, physical flows, and NPDSB indices after solving problem (\ref{oldattackmodel})-(\ref{newattackmodel_1}) for line $251$ with $\alpha = 10 \%$, $\alpha^{startpoint}_{251, min} = 0.0686$, and three different $d$s.}
\begin{tabular}{|c|c|c|c|c|c|} \hline
\multirow{2}{*}{\makecell{\textbf{Line}}} & \multirow{2}{*}{\makecell{\textbf{d}}} &  \textbf{Control room} & \textbf{Physical} &  \multirow{2}{*}{\makecell{\textbf{NPDSB}}} & \textbf{Flow limit}  \\
&  &  \textbf{flow (MW)} & \textbf{flow (MW)} & & \textbf{(MW)} \\
\hline
\multirow{3}{*}{\makecell{251}} & 285 &  -329.85 & -434.58 & 641 &\multirow{3}{*}{\makecell{387.34}} \\ \cline{2-5}
 & 485 &  -355.44 & -428.65 & 504 & \\ \cline{2-5}
 & 685 &  -317.16 & -387.347 & \textbf{374} &   \\
\hline
\end{tabular}
\label{line251thresh}
\end{table}

\subsection{Proposed Thresholds and RTLRTA Analysis}
\subsubsection{NPDSB$_{threshold}$ Analysis}
In this part, we demonstrated the functionality and accuracy of the proposed thresholds. To do so, we generated $2000$ scenarios of random LR attacks (red points) and $3000$ sets of random noise errors (blue points), including $2000$ Gaussian and $1000$ non-Gaussian noise errors. Then, we investigated if the proposed thresholds could detect the attack scenarios and differentiate them from random noise errors. Based on the determined thresholds for lines $169$ and $251$, every point with an NPDSB index larger than $350$ or $370$, respectively, was flagged.

\begin{table*}[bt]
\caption{The NPDSBs and physical overflows of all vulnerable transmission lines, considering one attack scenarios against lines $169$.}
\label{RTLRTAanalysisTable}
\centering
\scriptsize
\begin{tabular}{|c|c|c|c|c|c|c|c|c|c|}
\hline
\multirow{4}{*}{\textbf{\begin{tabular}[c]{@{}c@{}}Line No.\end{tabular}}} & \multirow{4}{*}{$169$} & \multirow{2}{*}{\textbf{P$_k^{max}$ (MW)}} & \multirow{2}{*}{$926.62$} & \multirow{4}{*}{\textbf{\begin{tabular}[c]{@{}c@{}}Attack \\ Scenario\end{tabular}}} & \multirow{4}{*}{$d =150$} & \textbf{NPDSB$_{52}^{169}$}  & $614$ & \multirow{4}{*}{\textbf{\begin{tabular}[c]{@{}c@{}}Physical \\ Overflow (\%)\end{tabular}}} & $21.6$  \\ \cline{7-8} \cline{10-10} 
                                                                              &                        &                                            &                           &                                                                                      &                           & \textbf{NPDSB$_{169}^{169}$} & $805$ &                                                                                             & $15.9$  \\ \cline{3-4} \cline{7-8} \cline{10-10} 
                                                                              &                        & \textbf{TNSB}                              & $1168$                    &                                                                                      &                           & \textbf{NPDSB$_{251}^{169}$} & $678$ &                                                                                             & $3.64$  \\ \cline{3-4} \cline{7-8} \cline{10-10} 
                                                                              &                        & \textbf{NPDSB$_{Threshold}$}                 & $350$                     &                                                                                      &                           & \textbf{NPDSB$_{264}^{169}$} & $332$ &                                                                                             & $13.18$ \\ \hline
\end{tabular}
\end{table*}

To achieve each random attack vector, we solved problem (\ref{oldattackmodel})-(\ref{newattackmodel_1}) for different $d$s, and each time $\alpha$ ($10 \%$) was multiplied to a random number between $0.52$ and $1$, where $0.052$ is the least $\alpha$ that causes $5 \%$ overflow on the target line ($\alpha^{5\%}_{169, min}$). We used this $\alpha$ to ensure that constraint (\ref{newattackmodel_1}) does not force many of the physical flows to be less than the target lines' continuous rating.

We generated the Gaussian noise errors' vectors from a Gaussian distribution with $\mu = 0$ and $\sigma = \alpha  L/3.1$ in such a way that the deviation at each bus was limited to $\alpha$ percent of the forecasted load at that bus in either direction. There was no change at zero injection buses, and the system's net load change was small. Moreover, we extracted the Cauchy noise errors' vectors from a Cauchy distribution with location $x_0 = 0$ and scale $\gamma = \alpha L/3.1$. All Cauchy noise errors were created and subjected to the same three constraints applied to the Gaussian noise errors' creation process.

\begin{figure}[!h]
 \centering
 \includegraphics[trim = 10mm 0mm 11.5mm 7mm, clip, width=7.528cm]{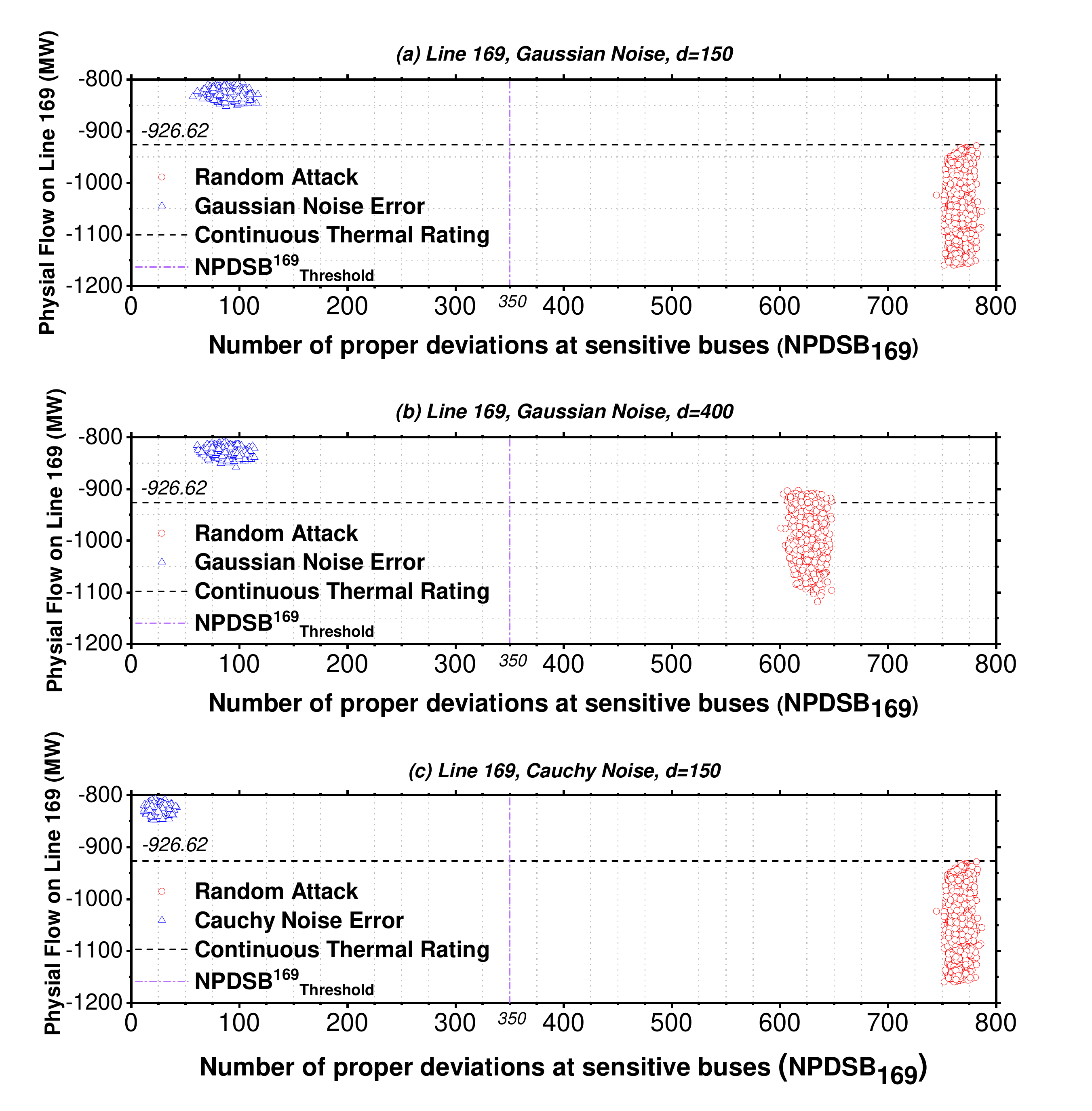}
 \caption{Physical effects of different scenarios of load deviations on line $169$ versus the associated
NPDSBs. Sub-figure (a) shows the comparison of physical effects of $1000$ random LR attacks, when
$d$ is $150$ in problem (\ref{oldattackmodel})-(\ref{newattackmodel_1}), with $1000$ sets of random Gaussian noise errors ($\mu = 0$; $\sigma = \alpha L/3.1$), sub-figure (b) shows the same comparison as the comparison in sub-figure (a), but considering $d$
equals to $400$, and sub-figure (c) shows the comparison between $1000$ sets of random LR attacks in
sub-figure (a) with $1000$ sets of Cauchy noise errors ($x_0 = 0$; $\gamma = \alpha L/3.1$).}
 \label{line169_attacknoise}
\end{figure}

\begin{figure}[!h]
 \centering
 \includegraphics[trim = 10mm 0mm 30mm 5mm, clip, width=7.38cm]{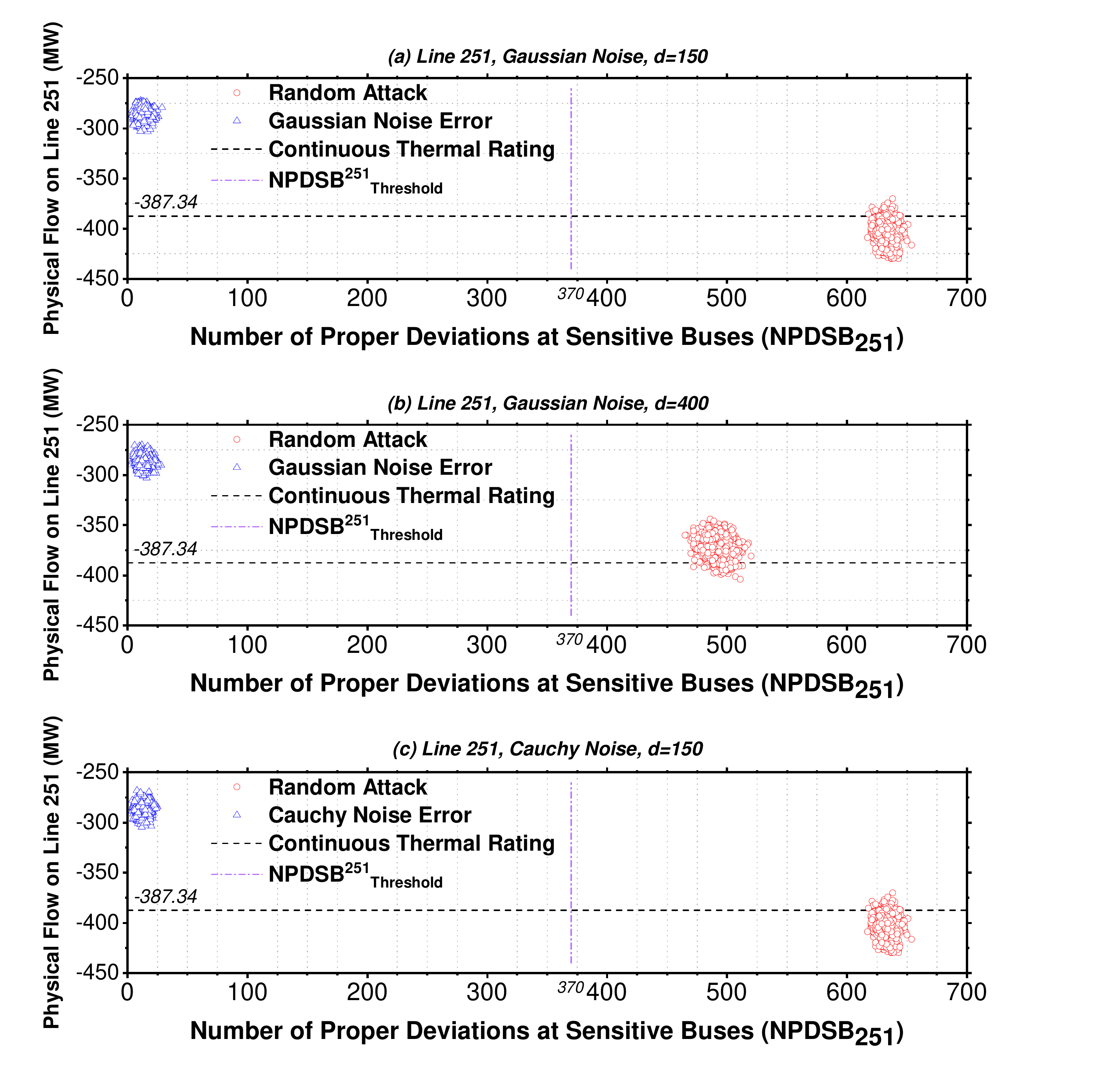}
 \caption{Physical effects of different scenarios of load deviations on line $251$ versus the associated
NPDSBs. Sub-figure (a) shows the comparison of physical effects of $1000$ random LR attacks, when
$d$ is $150$ in problem (\ref{oldattackmodel})-(\ref{newattackmodel_1}), with $1000$ sets of random Gaussian noise errors ($\mu = 0$; $\sigma = \alpha L/3.1$), sub-figure (b) shows the same comparison as the comparison in sub-figure (a), but considering $d$
equals to $400$, and sub-figure (c) shows the comparison between $1000$ sets of random LR attacks in
sub-figure (a) with $1000$ sets of Cauchy noise errors ($x_0 = 0$; $\gamma = \alpha L/3.1$).}
 \label{line251_attacknoise}
\end{figure}

As illustrated in Fig. \ref{line169_attacknoise} and Fig. \ref{line251_attacknoise}, the proposed thresholds worked perfectly and accurately. They successfully differentiated all random attack scenarios from noise errors for both lines. According to the results, some attack scenarios for both lines had insufficient energy to cause an overflow on the target line (red points above the lines related to continuous thermal ratings). It is because some of the $d$ randomly selected buses with zero deviations were among the most sensitive buses, which reduced the attack's energy. Although these scenarios were not successful in causing an overflow on the target line, our proposed method flagged them as a malicious movement since their NPDSB indices were larger than the pre-determined thresholds. We generated these scenarios (attack with no damage) to show our method’s capability, while this may not be the case in reality.

\begin{table*}[bt]
\scriptsize
\centering
\caption{Active and binding PLFSCs in the CPSCED, objective values of the SCED and CPSCED, and the physical flows when different scenarios of LR attacks ($d=150$ and $d=400$) contaminated the loads by targeting both transmission lines $169$ and $251$.}
\label{CPSCED_Table}
\begin{tabular}{|c|c|c|c|c|c|c|c|}
\hline
\textbf{Line No.}      & \textbf{Case No.} & \textbf{\begin{tabular}[c]{@{}c@{}}Activated\\ PLFSC \\ (index)\end{tabular}} & \textbf{\begin{tabular}[c]{@{}c@{}}Binding \\ PLFSC \\ (index)\end{tabular}} & \textbf{\begin{tabular}[c]{@{}c@{}}SCED\\ Cost (M\$)\end{tabular}} & \textbf{\begin{tabular}[c]{@{}c@{}}CPSCED\\ Cost (M\$)\end{tabular}} & \textbf{\begin{tabular}[c]{@{}c@{}}Target Line Physical \\  Flow after \\ SCED (MW)\end{tabular}} & \textbf{\begin{tabular}[c]{@{}c@{}}Target Line Physical \\  Flow after \\ CPSCED (MW)\end{tabular}} \\ \hline
\multirow{2}{*}{$169$} & $1$, $d=150$      & $52-169-264$                                                                  & $52$                                                                         & $1.79$                                                             & $1.83$                                                               & $-1011.7$                                                                                 & $-699.1$                                                                                    \\ \cline{2-8} 
                       & $2$, $d=400$      & \begin{tabular}[c]{@{}c@{}}$52-169-251- 264$\end{tabular}                 & $52 - 264$                                                                   & $1.78$                                                             & $1.82$                                                               & $-1039.4$                                                                                 & $-749.5$                                                                                    \\ \hline
\multirow{2}{*}{$251$} & $1$, $d=150$      & \begin{tabular}[c]{@{}c@{}}$52-169-251- 264$\end{tabular}                 & $52 - 264$                                                                   & $1.77$                                                             & $1.81$                                                               & $-409.1$                                                                                  & $-261$                                                                                      \\ \cline{2-8} 
                       & $2$, $d=400$      & $52-169-264$                                                                  & $52 - 264$                                                                   & $1.78$                                                             & $1.82$                                                               & $-383.9$                                                                                  & $-42.3$                                                                                     \\ \hline
\end{tabular}\end{table*}
\subsubsection{The RTLRTA Tool Analysis}
In this part, we evaluated the functionality of the RTLRTA tool. To do so, we ran Algorithm \ref{RTLRTANPDSB} when there was an LR attack scenario against target line $169$.
Hence, we generated a random LR attack against line $169$ by solving problem (\ref{oldattackmodel})-(\ref{newattackmodel_1}). We tested whether the RTLRTA tool can find all affected transmission assets or not. We calculated the NPDSB indices associated with the generated attack vector for other vulnerable transmission assets and compared each NPDSB index with its associated threshold to see if the transmission line was affected. Likewise, we added the malicious deviations related to this attack vector to the loads and provided the overflows' percentages on vulnerable transmission lines, which confirms the decision made by the RTLRTA tool.
As shown in Table \ref{RTLRTAanalysisTable}, we calculated the NPDSB indices for all four vulnerable transmission lines, which resulted in $614$, $805$, $678$, and $332$ for lines $52$, $169$, $251$, and $264$, respectively. Based on the thresholds, all NPDSB indices indicated a malicious movement and notified the system operator that the current loads were maliciously contaminated. Moreover, the overflows' percentages illustrate that this random attack, which was created to target line $169$, caused overflows on other vulnerable lines. As mentioned before, in this case, we took line $169$ as the primary target to estimate the actual loads since the overflow on this line ($21.6 \%$) is more significant than overflows on other vulnerable lines, which are $15.9 \%$, $3.64 \%$, and $13.18 \%$ for lines $52$, $251$, and $264$, respectively.

Algorithm \ref{RTLRTANPDSB} is fast enough to add to the existing EMSs; it was coded in JAVA and took about $50$ milliseconds to run for each line on an Intel(R) Xeon(R) CPU with 48 GB of RAM.
\begin{comment}
There are only four lines on which attackers can cause overflows for this specific test case, considering $\alpha$ at most $10 \%$. Hence, the operator needs to run Algorithm \ref{RTLRTANPDSB} only for these four vulnerable lines, which means that the RTLRTA tool takes around $200$ milliseconds to find whether the current set of loads has been maliciously contaminated or not.
\end{comment}
Likewise, since this problem’s nature makes each run of the RTLRTA tool for each asset independent from the others, we can parallelize the RTLRTA tool's running processes.

\subsection{CPSCED Analysis}
In this section, we evaluated the CPSCED's ability to provide secured dispatch points in the presence of LR attacks. To do so, we created different random LR attacks against transmission lines $169$ and $251$, and for each case, in the RTLRTA, we found the list of assets whose associated PLFSCs have to be activated in the CPSCED. Finally, we compared the operation costs of the SCED and CPSCED in the presence of LR attacks.
The proposed CPSCED has one more set of constraints than the SCED, which means that its result could be equal or worse from the economic point of view.  In other words, if any of the added PLFSCs binds, the CPSCED results in a higher operating cost than the operation cost of the SCED. 

Table \ref{CPSCED_Table} shows two different attack scenarios ($d=150$ and $d=400$) for each target transmission line ($169$ and $251$). As shown, there were binding PLFSCs for all four scenarios, which justifies that the CPSCED resulted in higher operating costs than the SCED. This is the cost of making the system secure against identified LR attacks. For instance, for case $1$ against line $169$, the PLFSC associated with line $52$ was the binding constraint out of the three activated PLFSCs, and resulted in higher operating cost in the CPSCED ($\$$ $1.83M$) than the SCED ($\$$ $1.79M$). As another example, consider the second case for line $251$, where the generated random attack was not successful in causing an overflow on this line. Still, it was successful on other lines ($52$, $169$, and $264$). The physical flow on line $251$ confirms the RTLRTA tool's decision, which did not activate this line's PLFSC in the CPSCED. This attack scenario caused overflows on the other three vulnerable lines ($52$, $169$, and $264$), and the RTLRTA successfully activated those PLFSCs in the CPSCED. Moreover, only two of these activated PLFSCs ($52$ and $264$) were binding in the CPSCED, which resulted in a higher operating cost in the CPSCED ($\$$ $1.82M$) than the SECD ($\$$ $1.78M$).

Considering the SCED's dispatch points in Table \ref{CPSCED_Table}, all the target lines’ actual physical flows were more than their thermal ratings (except for the second scenario against line $251$, in which we had other overloaded lines). On the other hand, considering the CPSCED dispatch points, there is no overflow in the system (neither on the target line nor on other vulnerable lines). 
\begin{comment}
For instance, in the second case for line $169$, the dispatch points from the SCED resulted in $-1039.4$ MW physical flow on the target line ($12.1 \%$ overflow), while the dispatch points from the CPSCED resulted in $-749.5$ MW flow on this line, which is within the acceptable range.
\end{comment}

\begin{comment}
The last two columns of Table \ref{CPSCED_Table} show the physical line flows on the target lines considering the dispatch points from both SCED and CPSCED. As shown, considering the SCED dispatch points, all the target lines' actual physical flows were more than the thermal ratings (except for the second scenario against line $251$. Although the second attack scenario against line $251$ did not cause any overflow on the target line (due to the random process), it caused overflows on other lines. 
On the other hand, considering the CPSCED dispatch points, there is no overflow in the system, which implies that an EMS equipped with our proposed approach can flag LR attacks and run the CPSCED, which provides secure dispatch points. 
\end{comment}

\begin{comment}
\textcolor{red}{In the second case for line $169$, the attack could cause $12.1 \%$ overflow. However, an EMS that is equipped with our proposed approach flags this attack and runs the CPSCED instead of the SCED, which results in no physical overflow in the system, as shown in Table \ref{CPSCED_Table}, -$749.5$ MW would be the flow on the target line.}
\end{comment}

% === IV. Conclusion Remarks ========================================
\section{Conclusion Remarks}
In the first part of this paper, we leveraged the NPDSB index to develop a real-time detection tool, which could be used in real-world practice with minimal disruptions and changes in the existing EMSs’ infrastructure. We made the RTLRTA tool in such a way that it could detect possible random LR attacks. To do so, first, we modeled the optimization problem (\ref{oldattackmodel})-(\ref{newattackmodel_1}) to generate random LR attacks. Next, we determined the NPDSB index's threshold for each vulnerable asset so that it can flag even the weakest random LR attack against that asset. Finally, we created $2000$ scenarios of random LR attacks and $3000$ Gaussian and non-Gaussian sets of noise errors to evaluate the determined thresholds and RTLRTA tool's accuracy and functionality. The determined thresholds worked perfectly, so the RTLRTA tool successfully detected all attack scenarios, differentiated them from all noise errors, and found all affected transmission assets. 

Second, we proposed an approach to estimate the actual loads, in the post-attack stage, based on which we designed the PLFSCs. Then, we added the set of PLFSCs to the SCED problem to create the CPSCED model as a corrective action to respond to the identified LR attacks. The CPSCED gets the affected transmission assets' ID from the RTLRTA tool and activates the associated PLFSCs to provide secured dispatch points. We investigated the CPSCED's functionality and compared its results with the SCED's results for different LR attack scenarios. The CPSCED successfully provided secured dispatch points, with no violation, for all attack scenarios, while the dispatch points from the SCED caused flow violations. Therefore, the higher operating cost of the CPSCED comparing to the SCED is the cost of making the system secure against undetectable LR attacks, which the SCED is incapable of doing that.
\ifCLASSOPTIONcaptionsoff
  \newpage
\fi

\bibliographystyle{IEEEtran}
\bibliography{IEEEabrv,Bibliography}

%\end{IEEEbiography}


\end{document}